\documentclass[pdflatex,sn-mathphys,iicol]{sn-jnl}

\jyear{2021}%

\theoremstyle{thmstyleone}%
\theoremstyle{thmstyletwo}%

\theoremstyle{thmstylethree}%
\usepackage{subcaption}
\RequirePackage{color}\definecolor{RED}{rgb}{1,0,0}\definecolor{BLUE}{rgb}{0,0,1} 
\RequirePackage[stable]{footmisc} 
\DeclareOldFontCommand{\sf}{\normalfont\sffamily}{\mathsf} 
\providecommand{\DIFaddbegin}{} 
\providecommand{\DIFaddend}{} 
\providecommand{\DIFaddbeginFL}{} 
\providecommand{\DIFaddendFL}{} 
\RequirePackage{listings} 
\RequirePackage{color} 
\lstdefinelanguage{DIFcode}{ 
  moredelim=[il][\color{red}\scriptsize]{\%DIF\ <\ }, 
  moredelim=[il][\color{blue}\sffamily]{\%DIF\ >\ } 
} 
\lstdefinestyle{DIFverbatimstyle}{ 
	language=DIFcode, 
	basicstyle=\ttfamily, 
	columns=fullflexible, 
	keepspaces=true 
} 
\lstnewenvironment{DIFverbatim}{\lstset{style=DIFverbatimstyle}}{} 
\lstnewenvironment{DIFverbatim*}{\lstset{style=DIFverbatimstyle,showspaces=true}}{} 

\begin{document}

\title[Technical Note]{Centreline shock reflection for Supersonic Internal Flows in the \textit{non-Rankine-Hugoniot} zone:\\ Overexpanded Supersonic Microjets}


\author*[1]{\fnm{Justin Kin Jun} \sur{Hew}}\email{u7322062@anu.edu.au}

\author[2]{\fnm{Hideaki} \sur{Ogawa}}\email{hideaki.ogawa@aero.kyushu-u.ac.jp}

\author[1]{\fnm{Rod W.} \sur{Boswell}}\email{rod.boswell@anu.edu.au}

\affil*[1]{\orgdiv{Space Plasma Power and Propulsion (SP3) Laboratory, Department of Nuclear Physics \& Accelerator Applications}, \orgname{Research School of Physics, Australian National University}, \orgaddress{ \city{Canberra}, \postcode{2601}, \state{ACT}, \country{Australia}}}

\affil[2]{\orgdiv{Department of Aeronautics and Astronautics, Graduate School of Engineering}, \orgname{Kyushu University}, \orgaddress{\city{Fukuoka}, \postcode{819-0395}, \country{Japan}}}


\abstract{The viscous and rarefaction effects on centreline shock reflection occurring in an overexpanded axisymmetric microjet have been investigated numerically by means of a fully coupled pressure-based shock capturing scheme. Due to the low free-stream Reynolds number (Re $\approx 7 )$, the Navier-Stokes equations were coupled with slip velocity and temperature jump boundary conditions to account for rarefied gas effects in the Knudsen layer. It has been found that pronounced viscosity levels can cause a transition from a three-shock to a two-shock configuration, which is impermissible by inviscid theory. This provides novel evidence that supports recent observations for axisymmetric ring wedge intakes. Analysis of the von Neumann and detachment criteria indicates that the transition from Mach reflection to regular\DIFaddbegin -like \DIFaddend reflection is analogous to the dual-solution domain transition for planar shocks. In addition, prediction of the longitudinal curvature of the incident shock has been conducted from a mathematical standpoint.}




\maketitle
\section{Introduction}\label{sec1}

It is well known that for wedge-induced planar shock reflection, both Mach reflection (MR) and regular reflection (RR) structures are possible. Moreover, the transition and hysteresis phenomenon of stable RR~$\rightarrow$ MR due to wedge angle variation, $\theta_w$ occurs abruptly, whereas the transition of stable MR $\rightarrow$ RR is a smooth transformation wherein the Mach stem height gradually approaches null \cite{Hornung1979}. This mechanism was investigated analytically by Hornung \& Robinson \cite{hornung1982transition} and Li \& Ben Dor~\cite{Li1998}, who showed that these cases corresponded to transitions occurring at the von Neumann condition for the steady flow regime. Thus, all the analytical, numerical and experimental studies so far clearly delineate the two possibilities for shock wave reflection in the planar case. 

For axisymmetric flowfields however, Rylov~\cite{rylov1990impossibility} proved that as two leading $C^{-}$ characteristic lines converge to the symmetry axis, they would always intersect before reaching the centreline, resulting in a limiting characteristic that steepens \DIFaddbegin the downstream portion of the incident shock \DIFaddend into an MR structure, precluding axisymmetric RR in theory. The dominant flow features consequently consist of an initially longitudinally curved incident shock, followed by a Mach stem slightly concave to the incoming flow, a reflected shock, slip stream, and a subsonic pocket initially downstream of the Mach stem called the \textit{Mach stream}~\cite{shoesmith2021modelling}. Additionally, further exploratory works by Hornung~\cite{hornung2000oblique} and Hornung \& Schwendeman~\cite{hornung2001oblique} have shown that other unique structures are possible such as the vortical Mach reflection (vMR) and the bulging Mach reflection with two triple points (DV).

The impossibility of RR in axisymmetric inviscid flow as proven much earlier was numerically observed by M\"older et al.~\cite{molder1997focusing}, Timofeev et al.~\cite{timofeev2001shock}, and Barkhudarov et al.~\cite{barkhudarov1991mach} by employing Euler codes for steady and unsteady shock reflection. They found that as the weak family shock wave steepened along the symmetry axis until the shock angle exceeded the von Neumann angle, the reflection structure would eventually bifurcate and form a three-shock configuration resembling MR.

However, contradicting results were obtained in experiments conducted by M\"older et al.~\cite{molder1997focusing} and Skews et al.~\cite{skews2002experiment} utilising flow visualisation techniques, which showed instead a two-shock configuration \DIFaddbegin (2SC)\DIFaddend, where the Mach stem \DIFaddbegin (disc) \DIFaddend did not appear. Hornung \cite{hornung2000oblique}, M\"older et al.~\cite{molder1997focusing}, Ogawa et al.~\cite{ogawa2013numerical,ogawa2014nhereumerical,ogawa2020centreline}, Shoesmith et al.~\cite{shoesmith2017shock} and Ogawa \& M\"older \cite{ogawa2014numerical} therefore suggested that under these conditions, as the longitudinally curved incident conical shock steepened toward the symmetry axis, it would reflect with either no Mach stem or a Mach stem too small to be observed. Additionally, M\"older et al.~\cite{molder1997focusing} and Shoev \& Ogawa~\cite{shoev2019numerical} attributed this result to the presence of Sternberg's non-\textit{Rankine-Hugoniot} zone \cite{sternberg1959triple} of highly viscous/rarefied axisymmetric flow.

Recently, Shoev \& Ogawa \cite{shoev2019numerical} numerically via  \DIFaddbegin Direct Simulation Monte Carlo (DSMC) \DIFaddend and \DIFaddbegin Navier-Stokes \DIFaddend no-slip approaches observed this  MR~$\rightarrow$~\DIFaddbegin 2SC \DIFaddend transition in an axisymmetric ring wedge intake for extremely low  Reynolds number (Re\DIFaddbegin $_{L} \approx 6.17$), where $L$ is the Mach disc radius from inviscid simulation\DIFaddend . Although not explicitly stated by the authors, their results revealed a gradual transition from stable MR to RR that is similar to that observed in the planar case. To our knowledge, this was the first such numerical study that demonstrated this possibility.

While a number of investigations have been conducted on the weak conical shock reflection on the symmetry axis, complete understanding of the phenomenon is yet to be achieved. Moreover, all the studies so far have been concerned with the axisymmetric ring wedge intake case for the purposes of fundamental understanding of the centreline shock reflection structure in the highly viscous regime, and no consideration has been placed on shock-induced separation as a result of adverse pressure gradient flows. Thus, to the best of our knowledge, this paper presents the first finding of the ~MR$~\rightarrow$~\DIFaddbegin 2SC \DIFaddend centreline shock transition for an axially symmetric overexpanded microjet.

\section{Computational Domain}
\begin{figure}
    \centering
    \includegraphics[trim = 4cm 1cm 3cm 10cm, width = \linewidth]{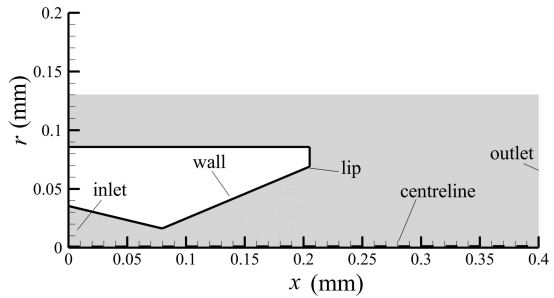}
    \caption{Schematic of the geometry of the micronozzle along with the far-field boundary. The dimensions are identical to Liu et al.~\cite{liu2006study}.}
    \label{fig:micronozzle}
\end{figure}
Figure~\ref{fig:micronozzle} displays the computational domain used in the present study, with all dimensions of the axisymmetric micronozzle being identical to the geometry employed by Liu et al.~\cite{liu2006study}, except an entrainment region that is also added for visualisation of the overexpanded jet plume.  

Grid generation is performed with unstructured quadrilateral cells for adaptive mesh refinement (AMR) based on density gradient as discussed in Section \ref{sec:3}. A symmetry axis boundary condition is also applied at the centreline due to the symmetric nature of the problem. 

Uniform inflow enters from the nozzle inlet, with a constant pressure of $P_i = 3 \times 10^5$~Pa, while the back pressure at the outlet is varied in small increments to alter the nozzle pressure ratio ($NPR$).  

\section{Numerical Framework}\label{sec:3}
The commercial code Fluent is used in the present study, with the setup and implementation methods as described below. 
A fully coupled pressure based solver based on collocated variable arrangement is used to solve the Navier-Stokes equations, which are discretised in conservative form via the finite volume method. Convective and diffusive fluxes are spatially discretised via van Leer's third-order TVD MUSCL scheme, with gradients computed using the Green-Gauss cell-centred interpolation approach for face values. Barth-Jespersen flux limiters are imposed to prevent spurious oscillations, and the linearised equations are implicitly time-advanced via the Lower-Upper Symmetric Gauss Seidel algorithm (LU-SGS) until steady-state convergence. Adaptive mesh refinement (AMR) techniques \cite{berger1989local} based on a dynamic density gradient criterion are also employed for all computations. 

The gas is assumed to be an ideal Newtonian fluid (air) with specific heat ratio $\gamma = 1.4$, dynamic viscosity is computed with Sutherland's law using the same parameters as described in Shoev \& Ogawa \cite{shoev2019numerical}.  \DIFaddbegin The Reynolds number used in this study is defined based on nozzle-exit and ambient parameters as:
\begin{equation}
\textrm{Re} = \frac{\rho_{\infty} a_{\infty } r_e}{\mu} = \frac{\sqrt{\gamma}}{\mu}\frac{p_{\infty} r_e}{\sqrt{R T_{\infty}}} 
\end{equation}
where $\rho_{\infty}$, $a_{\infty}$ ,$T_{\infty}$ and $p_{\infty}$ are ambient density, acoustic speed, temperature and pressure, respectively. $\mu$ is the dynamic viscosity, $R$ is the specific gas constant and $r_e$ is the nozzle-exit radius.
\DIFaddend Moreover, as demonstrated by Zeitoun et al. \cite{zeitoun2009numerical}, Zeitoun \& Burtschell \cite{zeitoun2006navier} and numerous other rarefied SBLI studies \cite{babinsky2011shock}, rarefaction effects within the slip and early transition regime (Kn~$\approx~0.1$) can be accounted for by the imposition of slip velocity and temperature jump boundary conditions along the wall~\cite{kogan2013rarefied}. In particular, the Maxwell-Smoluchowski relations are utilised, given as:
\begin{equation}
u_{s}-u_{w}=\alpha_{u} \lambda\left(\frac{\partial u_{\tau}}{\partial y}\right)_{s} 
\end{equation}
\begin{equation}
T_{s}-T_{w}=\alpha_{T} \frac{\gamma}{\gamma-1} \frac{\lambda}{\mathrm{Pr}}\left(\frac{\partial T}{\partial y}\right)_{s}
\end{equation}
where Pr is the Prandtl number, $u_{\tau}$ is the tangential component of velocity, $\lambda$ is the collisional mean-free-path (MFP) and $T$ is the static temperature. $\alpha_u$ and $\alpha_T$ are the tangential momentum and thermal accommodation coefficients, respectively. The subscripts $s$ and $w$ denote slip and wall regions of the Knudsen layer, respectively. Following Zeitoun et al. \cite{zeitoun2009numerical}, $\alpha_u = 1.142$ and $\alpha_T = 0.5865$ are used.

Most of the computations, including the grid independence study (Section \ref{secgrid}) were performed on a 96-core (2 nodes) Intel Xeon Platinum 8274 (Cascade Lake) with 3.2 GHz CPUs and 192GB memory per node on Gadi at the National Computational Infrastructure (NCI). 

\section{Results and Discussion}\label{sec4}
\subsection{Grid independence study and flowfield analysis}\label{secgrid}
A grid resolution study is conducted for the main case wherein a clear MR $\rightarrow$~\DIFaddbegin 2SC \DIFaddend transition is observed. AMR is utilised to resolve the shock reflection structure across numerous cell stencils. While prior work ensured that the shocks were resolved with 10 computational cells inside it, the present study has gone well beyond that and resolve them with at least 50 computational cells. This is to ensure that our observations are not the result of grid resolution, as was the concern raised by Timofeev et al.~\cite{timofeev2001shock}, Shoev \& Ogawa \cite{shoev2019numerical} and Isakova et al.~\cite{isakova2012amplification}\DIFaddbegin , where the disappearance of the Mach disc was possibly attributed to a numerical artefact\DIFaddend . 
\begin{figure}[!h]
    \centering
    \includegraphics[width = \linewidth]{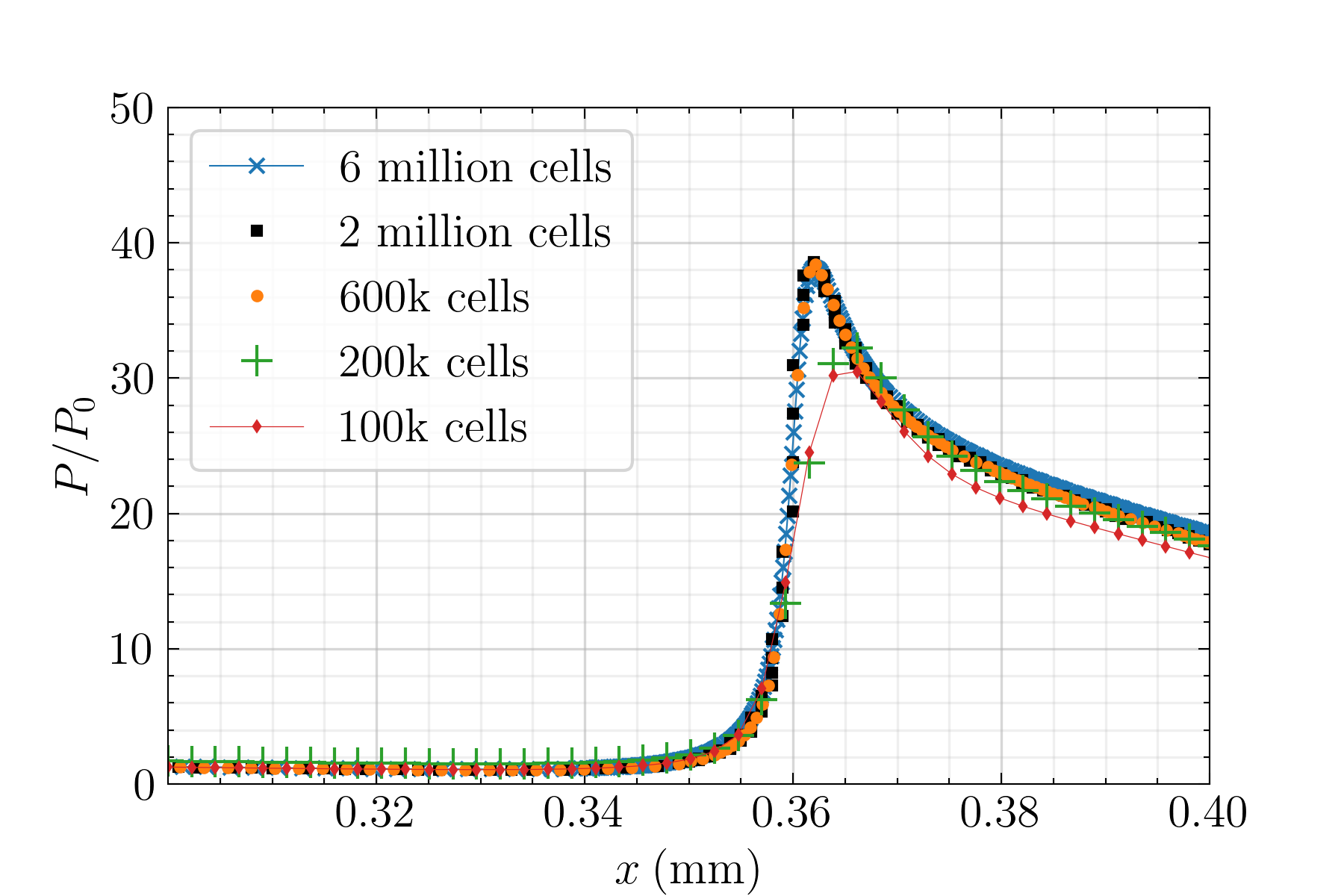}
    \caption{Non-dimensional static pressure distribution along the centreline for various grid densities at nozzle pressure ratio ($NPR$)~$\approx 55$, where \DIFaddbegin weak two-shock reflection (w-2SC) \DIFaddend is observed.}
    \label{fig:static}
\end{figure}
\DIFaddbegin 

\begin{figure}
    \centering
     \DIFaddbeginFL \includegraphics[trim = 4cm 1cm 3cm 4cm,width = \linewidth]{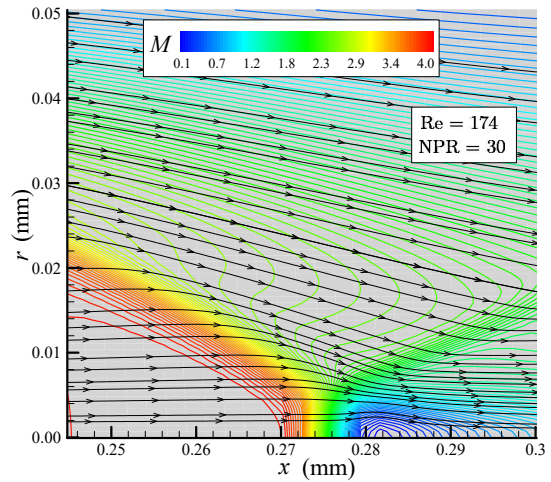}
    \DIFaddendFL \caption{Mach number contour displaying the Mach reflection configuration at Re \DIFaddbeginFL $ = 174$ \DIFaddendFL and $NPR = 30$. }
    \label{fig:machreflec}
\end{figure}

\begin{figure}
    \centering
     \DIFaddbeginFL \includegraphics[trim = 4cm 1cm 3cm 4cm, width = \linewidth]{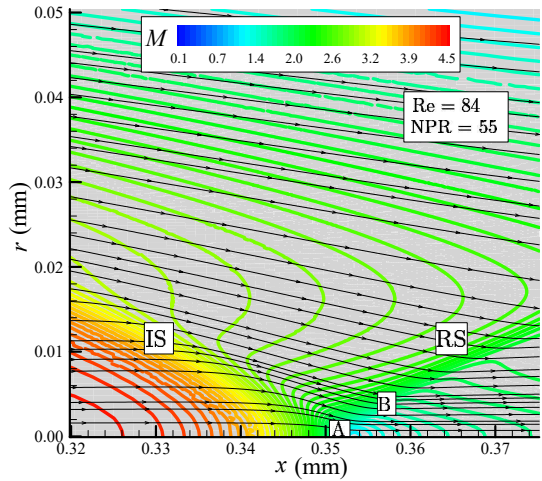}
    \DIFaddendFL \caption{Mach number contour of the \DIFaddbegin weak two-shock configuration (w-2SC) \DIFaddend configuration at Re $= 84$ , $NPR$ = 55.}
    \label{fig:regref}
\end{figure}

\DIFaddend With increasing the $NPR$, the Mach  \DIFaddbegin disc radius\DIFaddend , along with the {\it Mach stream}, decreases in size due to the weakening of the recompression shock. Moreover, Fig. \ref{fig:machreflec} shows that the streamlines downstream of the curved Mach stem are inclined upwards, indicating that it is an inverted Mach reflection (InMR), rather than direct. As the $NPR$ is further increased, a transition to weak \DIFaddbegin 2SC \DIFaddend occurs, \DIFaddbegin akin to weak planar \DIFaddend regular reflection (wRR), where the Mach stem vanishes and the downstream flow remains supersonic (Fig. \ref{fig:regref}). 

\begin{figure}
    \centering
    \includegraphics[width = \linewidth]{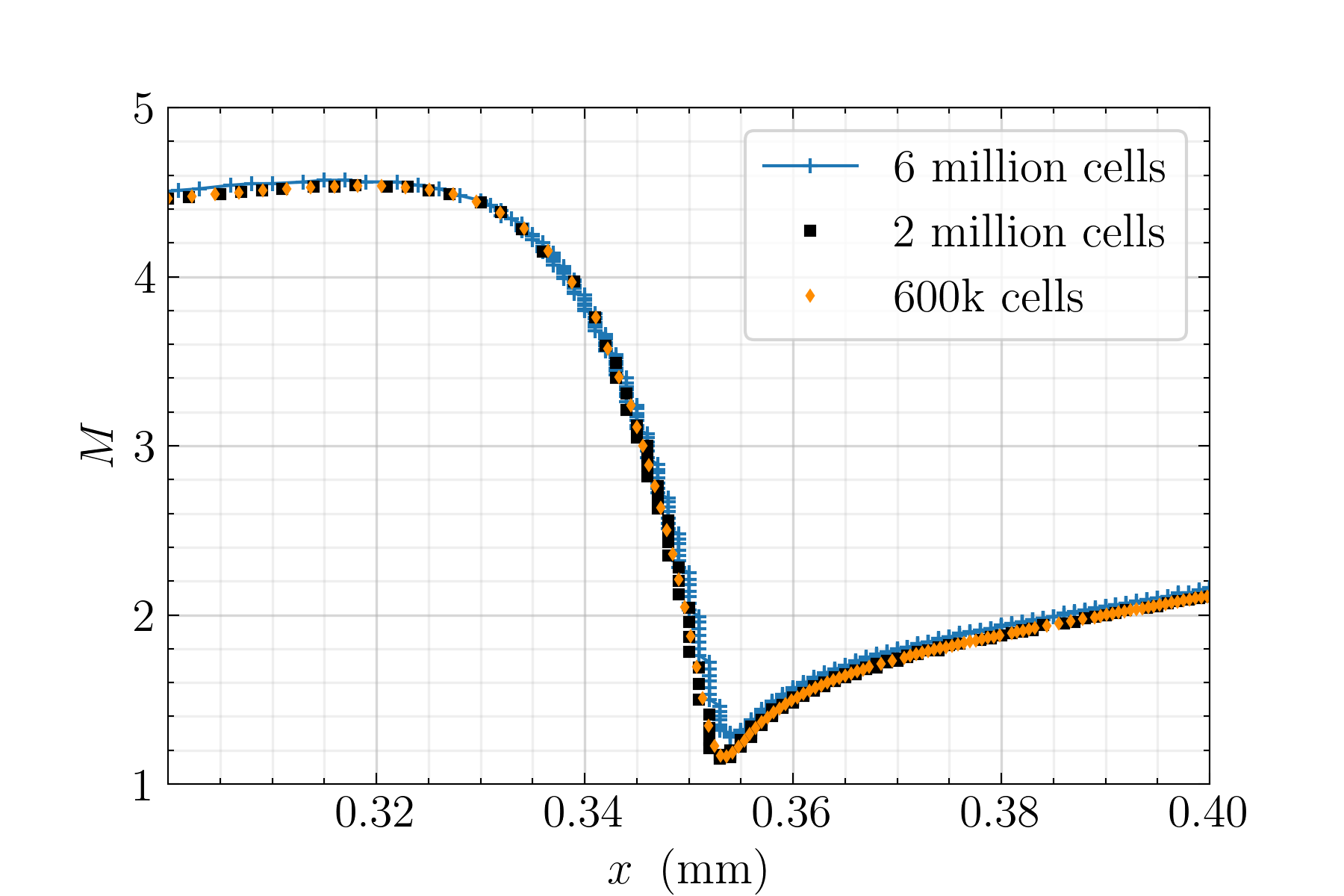}
    \caption{Centreline Mach number distribution for various grid densities at nozzle pressure ratio ($NPR$)~$\approx 55$, where \DIFaddbegin weak two-shock configuration \DIFaddend \DIFaddbegin 2SC \DIFaddend (w-2SC) is observed.}
    \label{fig:machno}
\end{figure}
\begin{figure}
    \centering
    \includegraphics[width = \linewidth]{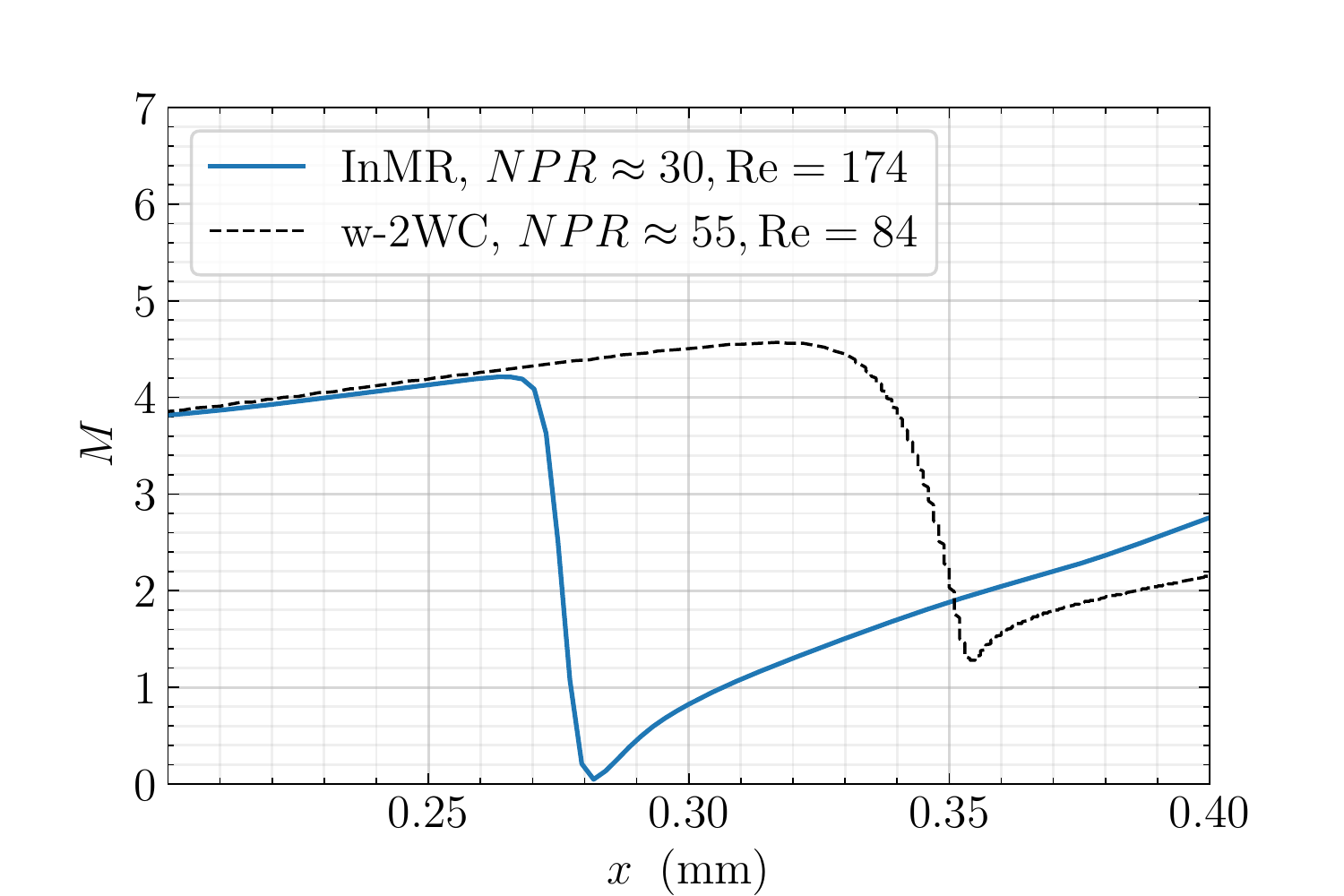}
    \caption{Centreline Mach number distribution of both the inverted Mach reflection (InMR) configuration in Fig.~\ref{fig:machreflec}, and the \DIFaddbegin weak two-shock configuration (w-2SC) \DIFaddend in Fig.~\ref{fig:regref}. The InMR has a post-shock Mach number that drops to subsonic levels due to the presence of a Mach stem, while the w-2SC does not.  }
    \label{fig:machplot}
\end{figure}
To demonstrate grid independence in the main case of \DIFaddbegin w-2SC \DIFaddend, the non-dimensional static pressure distribution across the centreline shock structure is plotted for 5 different grid densities (Fig. \ref{fig:static}). It demonstrates that the $100$k and $200$k cell cases do not suffice in terms of numerical resolution; but with denser grids, namely $600$k onwards, the results fast approach a converged distribution, so that the solution is independent of the grid resolution. The centreline Mach number distribution (Fig. \ref{fig:machno}) similarly demonstrates the converged behaviour. Both Figs. \ref{fig:machno} and \ref{fig:machplot} also clearly show that the post-shock Mach number of the weak \DIFaddbegin weak two-shock structure \DIFaddend never drops to unity or subsonic levels for all grid resolutions. This therefore indicates the absence of any Mach stem that would normally be present in axisymmetric internal shock reflection~\cite{rylov1990impossibility,isakova2012amplification}, consistent with~\cite{shoev2019numerical}. Therefore, the possibility of MR~$\rightarrow$~\DIFaddbegin w-2SC \DIFaddend is clearly elucidated in the grid independence study, where no Mach stem or subsonic {\it Mach stream} is formed due to pronounced viscous and rarefaction effects.

\begin{figure}[]
    \centering
    \begin{subfigure}[b]{\columnwidth}
        \centering
        \includegraphics[width=\linewidth]{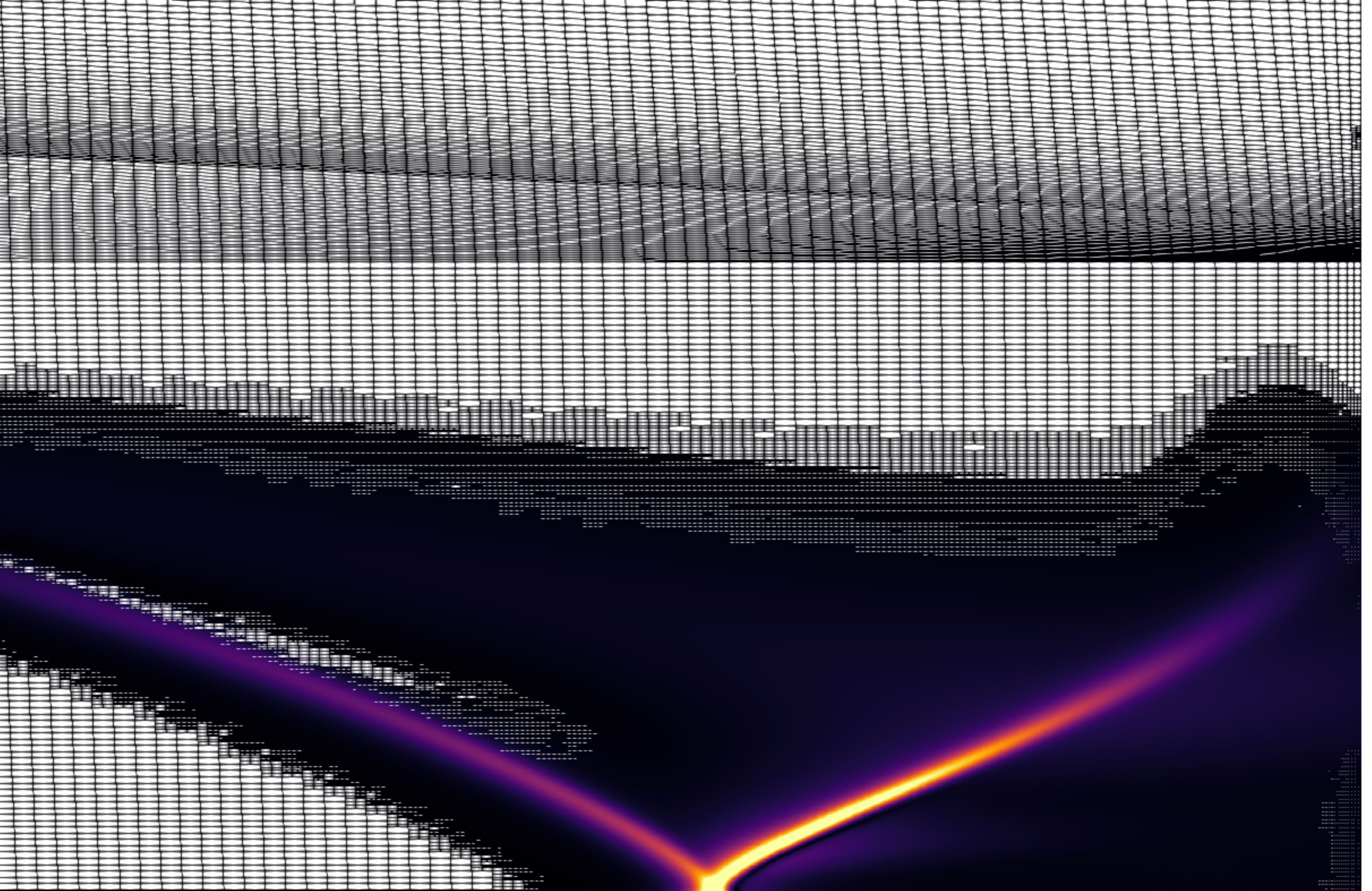}
        \caption{\label{fig:schlieren}}

    \vspace{0.1cm}
    \end{subfigure}
    \begin{subfigure}[b]{\columnwidth}
        \centering
        \includegraphics[width=\linewidth]{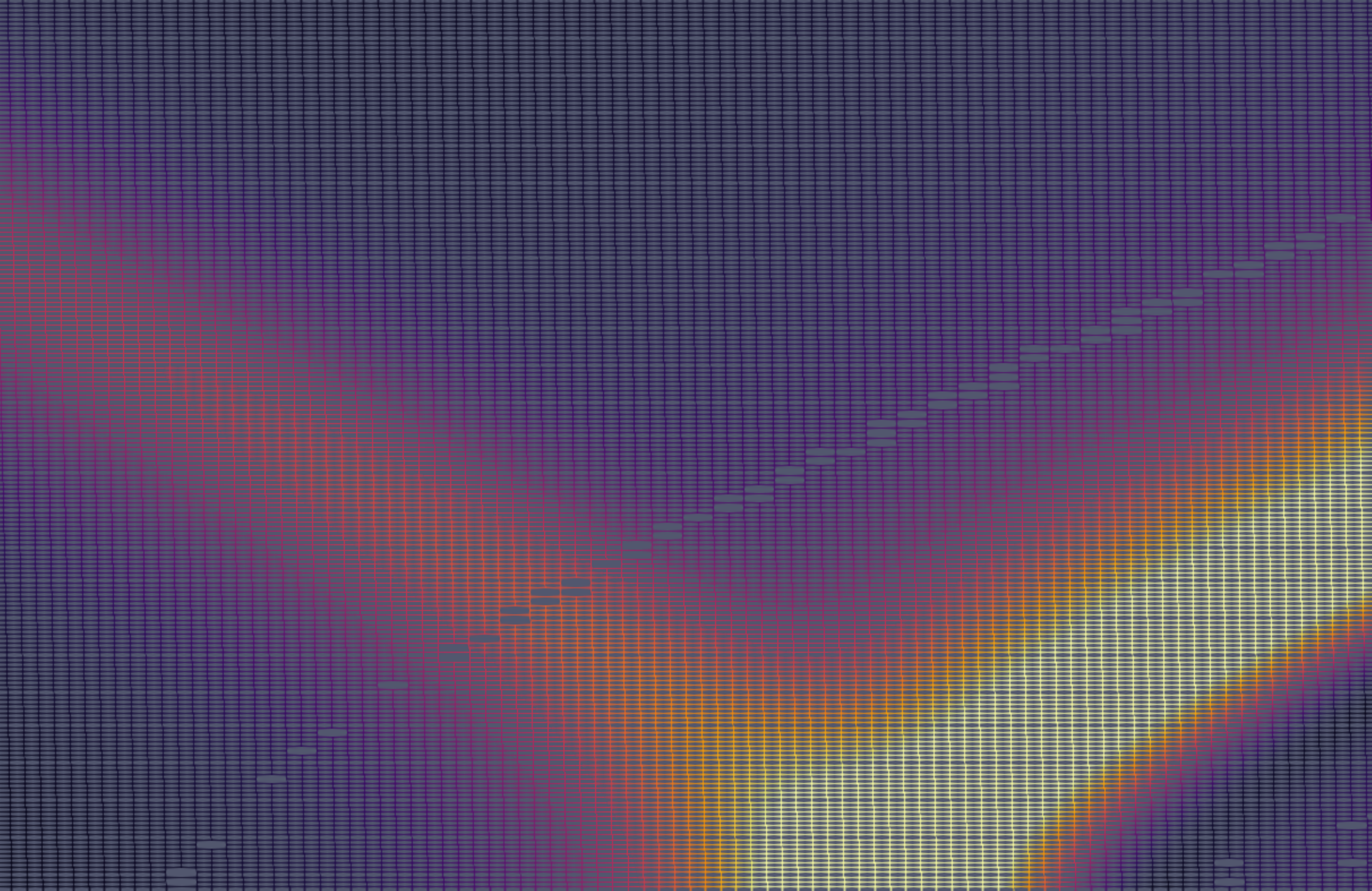}
        \caption{\label{fig:close}}

    \end{subfigure}
    \caption{(a) Numerical schlieren of the \DIFaddbegin weak two-shock configuration (w-2SC) \DIFaddend with 10 AMR levels of order 3 reaching $2250 \times 2250$ cells in total, at Re  \DIFaddbeginFL = 84\DIFaddendFL , $NPR$ = 55. (b) Close-up view of Fig. \ref{fig:schlieren} showing the grid density at the point of reflection on the reflecting surface for Re$_{\delta} =~7$, $NPR$ = 55. Almost $50$ cells are embedded transversely within the shock, well surpassing the resolution achieved in prior works demonstrating the possibility of MR~$\rightarrow$~\DIFaddbegin 2SC \DIFaddend transition in axisymmetric flow \cite{shoev2019numerical}, further validating its possibility.}
    \label{fig:fpr-tpr}
\end{figure}

A numerical schlieren of the \DIFaddbegin w-2SC \DIFaddend is also displayed in Fig.~\ref{fig:schlieren}, where 10 AMR levels have been imposed to reach a total cell count of about 6.3 million, densely resolving the shock structure. As seen in the close-up view  near the point of reflection (Fig.~\ref{fig:close}), more than $50$ cells are embedded transversely within the incident and reflected shocks, well surpassing the resolution achieved in prior works \cite{shoev2019numerical}, eliminating the possibility that the present observations might be due to insufficient grid resolution. 

It is also interesting to note that the MR~$\rightarrow$~\DIFaddbegin 2SC \DIFaddend transition occurs  \DIFaddbegin as \DIFaddend the Reynolds number  \DIFaddbegin decreases to 84 \DIFaddend here (Fig. \ref{fig:regref}), which is  \DIFaddbegin qualitatively consistent with \DIFaddend that observed by Shoev \& Ogawa \cite{shoev2019numerical}  in their ring wedge intake simulation. Their results similarly indicated that  \DIFaddbegin as the Reynolds number decreases\DIFaddend , a Mach reflection configuration  \DIFaddbegin (Fig. \ref{fig:machreflec}) \DIFaddend with a small but visible Mach stem \DIFaddbegin becomes a two-shock configuration, in contrast with inviscid theory\DIFaddend . Therefore, this suggests the possibility of a kind of limiting Reynolds number where such transitions can occur, so that these shock structures only appear in extremely viscous and rarefied flowfields. \DIFaddbegin This property is investigated further in Section~\ref{sec:reynolds}.
\DIFaddend 

\subsection{Shock polar and transition criteria analysis}

A unique feature that characterises an inward-turning conical shock structure is its longitudinal curvature~\cite{gounko2017patterns,shoev2019numerical}, which means that a shock polar cannot be defined uniquely based on two- or three-shock theory. 
Due to this characteristic, flow deflection angle only near the point of reflection is considered in this study. Additionally, the effects of viscosity and curvature are investigated by studying the longitudinal curvature near the point of reflection with the method adopted by Shoev \& Ogawa \cite{shoev2019numerical}, based on the computational results (Fig.~\ref{fig:regref}).

\begin{figure}[H]
    \centering
     \DIFaddbeginFL \includegraphics[width =\linewidth]{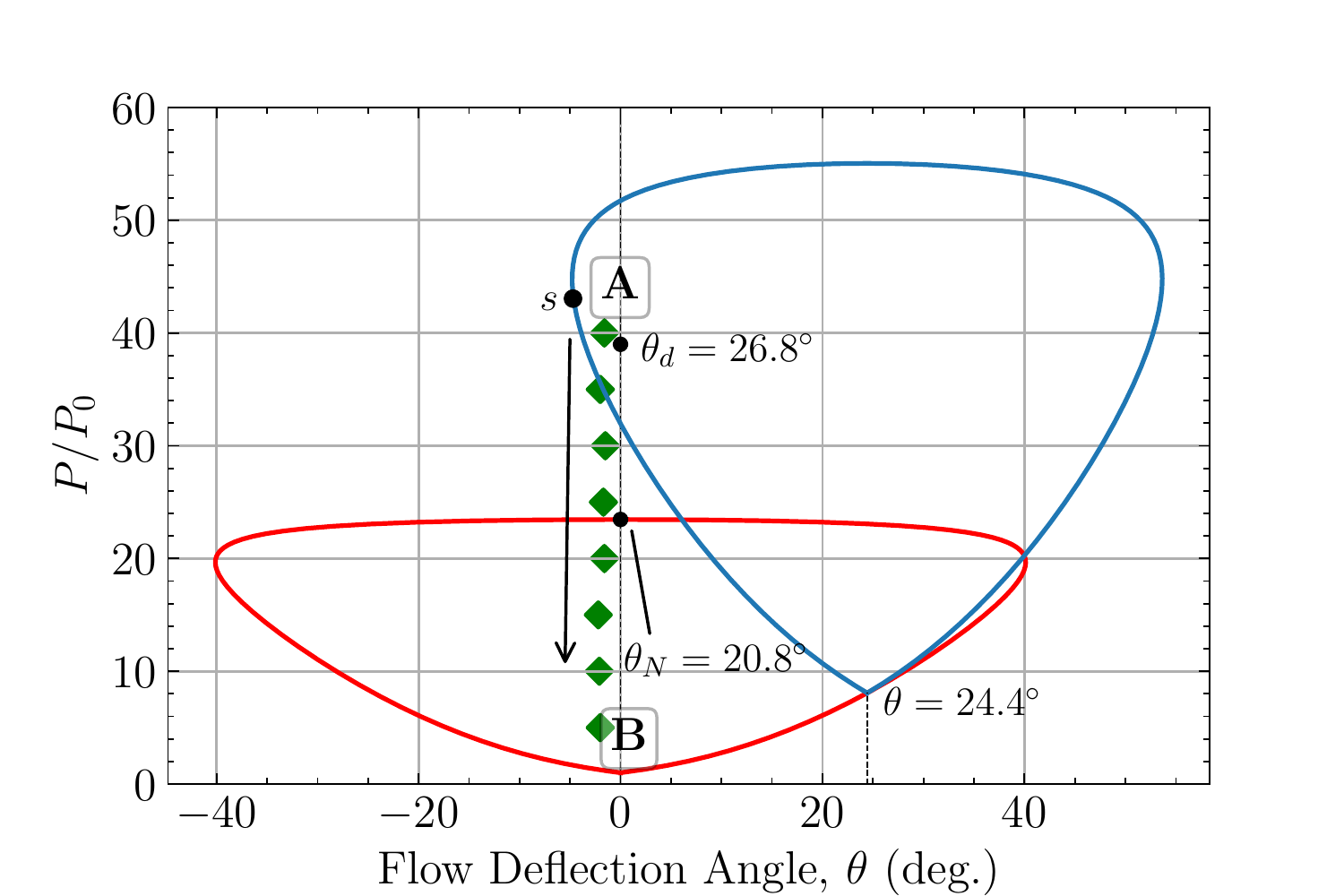}
    \DIFaddendFL \caption{Shock polar representation of \DIFaddbegin w-2SC/wRR \DIFaddend observed in the numerical simulations, where $\theta_1=~ 24.4^{\circ}$ and $M_0=4.5$, showing the locus of all possible flow states in the reflection process. The red shock polar is the incident polar (i-polar), while the blue is the reflected polar (r-polar). $\theta_d$ and $\theta_N$ are deflection angles corresponding to detachment and von Neumann criteria, respectively, and $s$ denotes the sonic criterion on the r-polar, which divides the weak and strong branches. The green markers illustrate the local flow change in the downstream vicinity of the point of reflection (A-B), as denoted in Fig.~\ref{fig:regref}. }
    \label{fig:shockpolar}
\end{figure}

\begin{figure}[H]
    \centering
    \includegraphics[width = \linewidth]{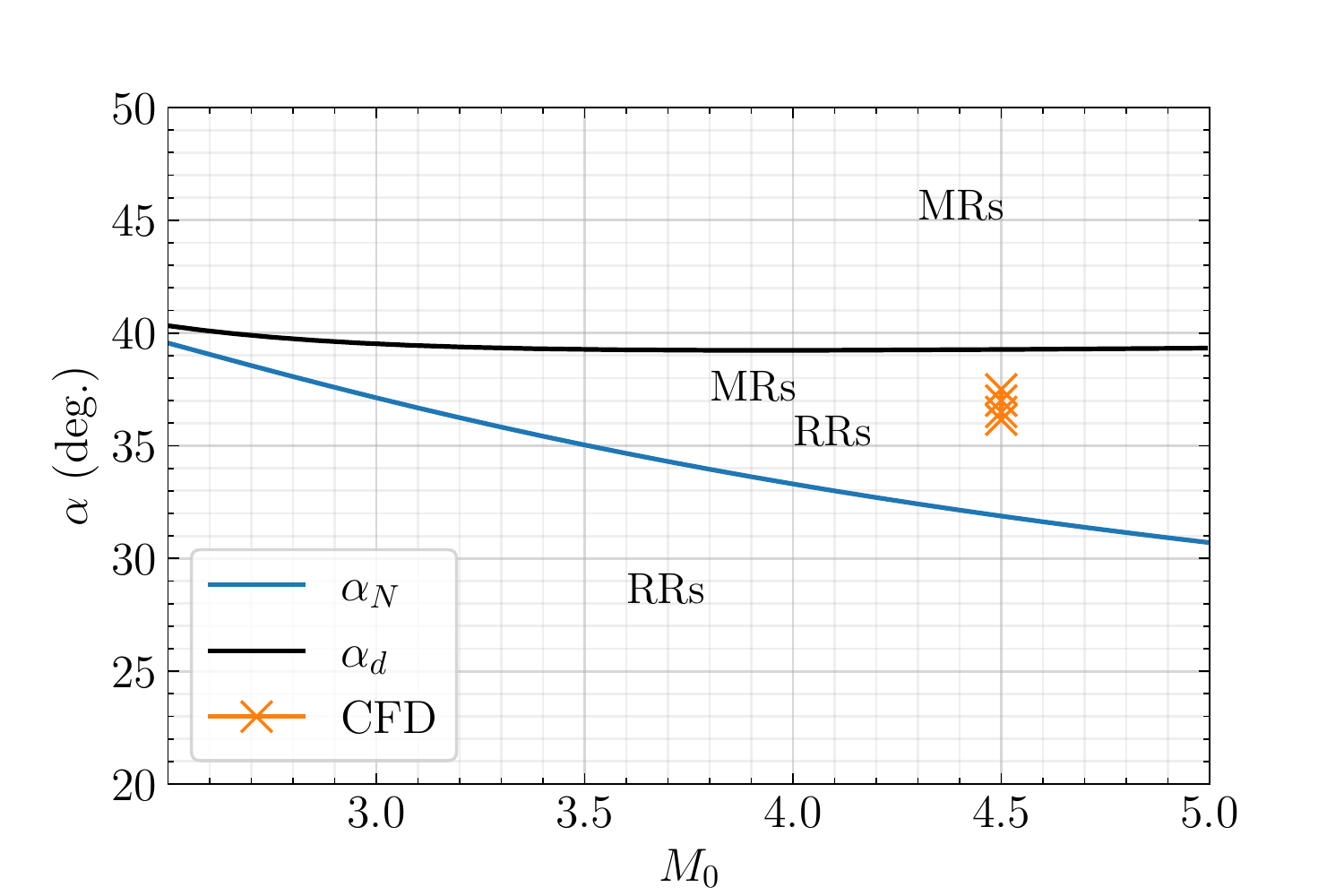}
    \caption{Shock reflection configurations in the ($M_0,\alpha$) plane and the transition criteria between them, where $\alpha_N$ is the von Neumann angle,  and $\alpha_d$ the detachment angle. The orange ($\times$) markers indicate that the incident curved shock angles from computational calculations (CFD) correspond to MR~$\rightarrow$ RR transition within the dual-solution domain for $M_0 = 4.5$. The curved incident shock angles in CFD are measured continuously from the triple point until the changes become negligibly small. }
    \label{fig:vonneumann}
\end{figure}

The incident (i) and reflected (r) polars (Fig. \ref{fig:shockpolar}) are plotted for $M_0 = 4.5$, with flow deflection angle $\theta_1 = 24.4^{\circ}$. It can be seen that the r-polar crosses the zero-deflection line well above the i-polar and also on its ``weak" solution portion, which indicates that it is either a wRR or MR-type configuration. Moreover, the flow deflection angle of $24.4^{\circ}$ corresponds to a transition occurring in between the von Neumann and detachment angles, as indicated in both Figs.~\ref{fig:shockpolar} and~\ref{fig:vonneumann}. The green markers denote the flow deflection angle across the regions $A$ and $B$ specified in Fig. \ref{fig:regref}, which has minimal flow deflections due to the absence of a curved Mach stem.
 \DIFaddbegin 

\DIFaddend As seen in Figs.~\ref{fig:shockpolar} and~\ref{fig:vonneumann}, the computational results signify that the MR $\rightarrow$ RR transition occurs within the dual-solution domain, \DIFaddbegin which implies an analogy with \DIFaddend the dual-solution domain transition \DIFaddbegin in\DIFaddend planar shock reflection \DIFaddbegin , at least based on the theoretical transition boundaries of two and three-shock theories. It indicates that three to two-shock transition will always occur above the von Neumann criterion \cite{molder1997focusing,hornung2000oblique}, in the region where both MR and RR are possible. Whether or not a similar hysteresis pattern between a three-shock and two-shock structures can also be observed in axisymmetric flows is still an unanswered question, and was not observed in the present work. 
Additionally, we also \DIFaddend note that Shoev \& Ogawa\DIFaddbegin ~\DIFaddend \cite{shoev2019numerical} similarly observed the transition from a stable three- to two- shock configuration within the dual-solution domain, although it was not explicitly mentioned by the authors. 
\DIFaddbegin \subsection{Limiting Reynolds number}\label{sec:reynolds}
Since it has been shown that enhanced viscosity levels can induce transition from a three-shock to a two-shock configuration, with no downstream subsonic flow (w-2SC by nomenclature), the possibility of a limiting Reynolds number where these structures can emerge becomes a natural question. Therefore, this phenomenon is investigated further by re-simulating the present cases, and changing only the wall boundary condition from a partial slip to a free-slip one. This ensures the removal of any boundary layer. Thus, there are no flow separation effects, which are known to accelerate transition in the planar regime \cite{shimshi2009viscous}, as the freestream Mach number is reduced. 
\begin{figure}[tbh]
    \centering
    \begin{subfigure}[b]{\columnwidth}
        \centering
        \includegraphics[trim = 1cm 2cm 1.5cm 2cm,width=\linewidth]{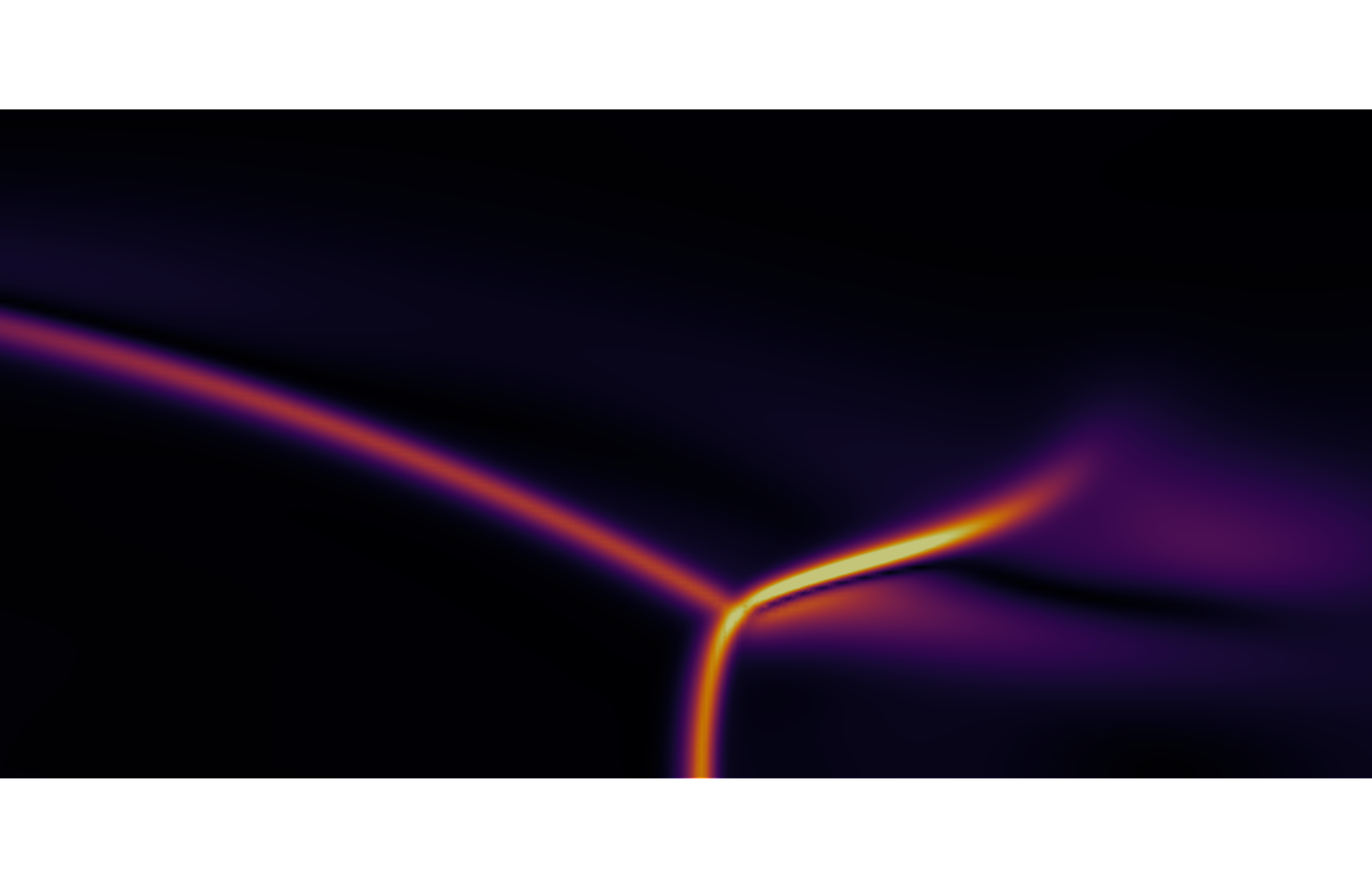}
        \caption{\label{fig:freeslip1}}
        \DIFaddendFL 

    \DIFaddbeginFL \vspace{0.1cm}
    \end{subfigure}
    \begin{subfigure}[b]{\columnwidth}
        \centering
        \includegraphics[trim = 1cm 2cm 1.5cm 2cm,width=\linewidth]{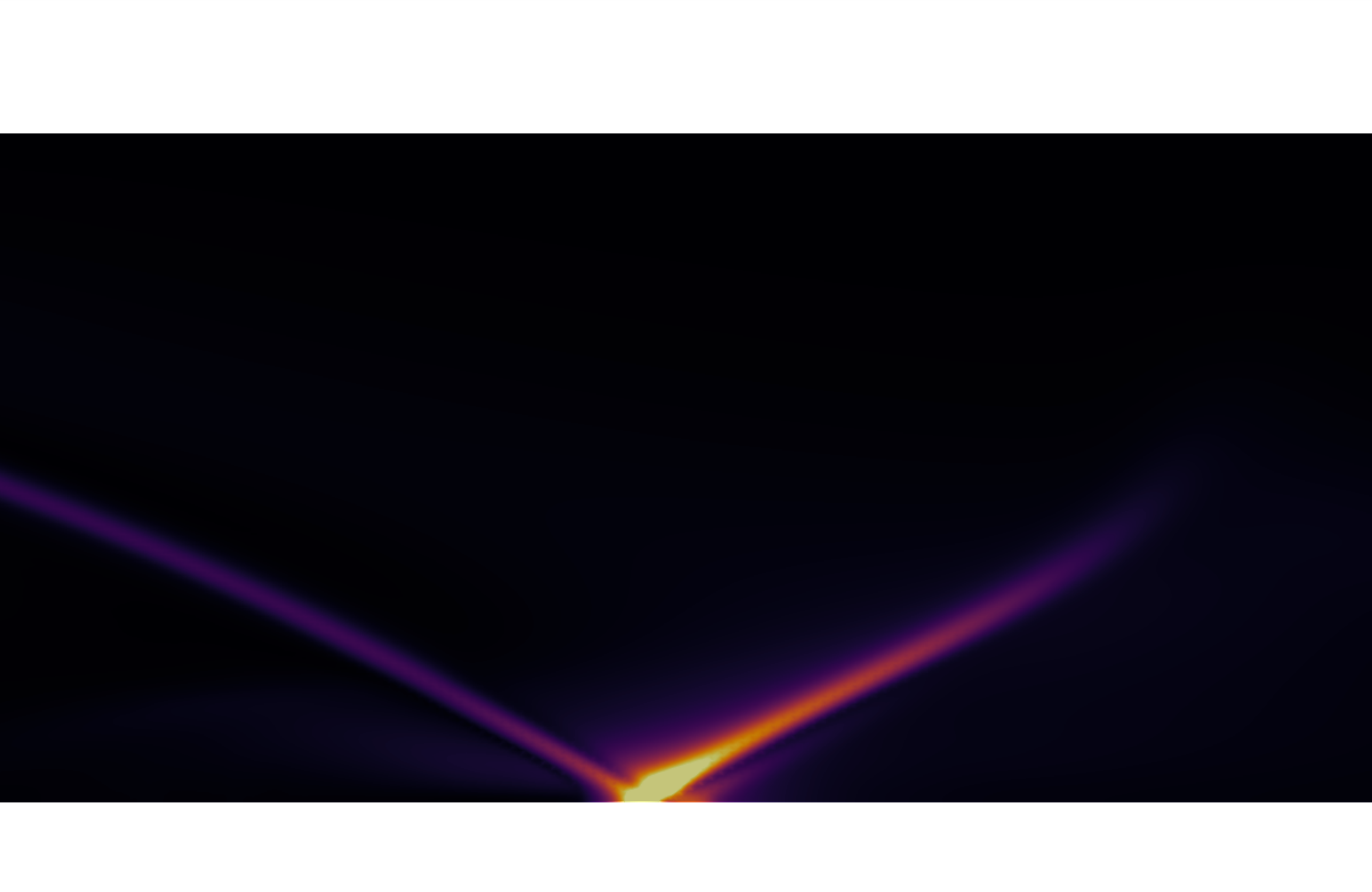}
        \caption{\label{fig:freeslip2}}

    \end{subfigure}
    \caption{Numerical schlieren images, (a) inverted Mach reflection (InMR) at Re~$\approx$ 92, $NPR$~=~60. (b) weak two-shock configuration (w-2SC) for Re~$ = 7.2$, $NPR \approx$ 600.}
    \label{fig:freeslip}
\end{figure}
Numerical schlieren images of the flowfields wherein a three- to two-shock transition occurs is shown in Fig.~\ref{fig:freeslip}, at $M_0 = 5.9$. A similar observation as in the planar flows is also made in the present results,  that the disappearance of the Mach disc now occurs at a higher NPR due to the larger inflow Mach number \cite{shimshi2009viscous}. This observation is consistent with the fact that there exist some similarities between the shock reflection transition patterns of planar and highly viscous axisymmetric flows. In this case, it is clearly elucidated that the transition pattern to a two-shock system is at a low Reynolds number of $\textrm{Re} \approx 7.2 $, while MR is found at high Re. Similar observations were also made in~Ref.~\cite{shoev2019numerical}. Thus, the results strongly suggest the possibility of a limiting or critical Reynolds number $\textrm{Re}_{\textrm{crit}}$ for Mach disc disappearance, and consequently the three to two shock transition.

\DIFaddend \subsection{Analogy to the Guderley's Singularity}
The \DIFaddbegin Guderley-Landau-Stanyukovich \DIFaddend implosion problem \cite{guderley1962theory} occurs when an incoming spherical or cylindrical shock steepens toward an axis of symmetry, wherein a singularity in the solution is found due to the presence of a limiting characteristic. Guderley \cite{guderley1962theory} fitted the shape of the incident shock with a family of similarity solutions in the $(r,t)$ plane, where $r$ is the radial coordinate and $t$ denotes time. Utilising Guderley's method, Hornung \& Schwendeman \cite{hornung2001oblique} considered axisymmetric shock reflection within a steady flowfield in an analogous manner, and confirmed the analogy via inviscid simulations for overexpanded supersonic jets. Shoev \& Ogawa \cite{shoev2019numerical} also examined the analogy for their axisymmetric ring wedge intake simulations in a viscous flowfield. Therefore, the same approach is employed in the present study to investigate its applicability for highly viscous overexpanded microjets. 

\begin{figure}[hbtp]
    \centering
     \DIFaddbeginFL \includegraphics[trim = 1cm 1cm 1cm 1.5cm, width = \linewidth]{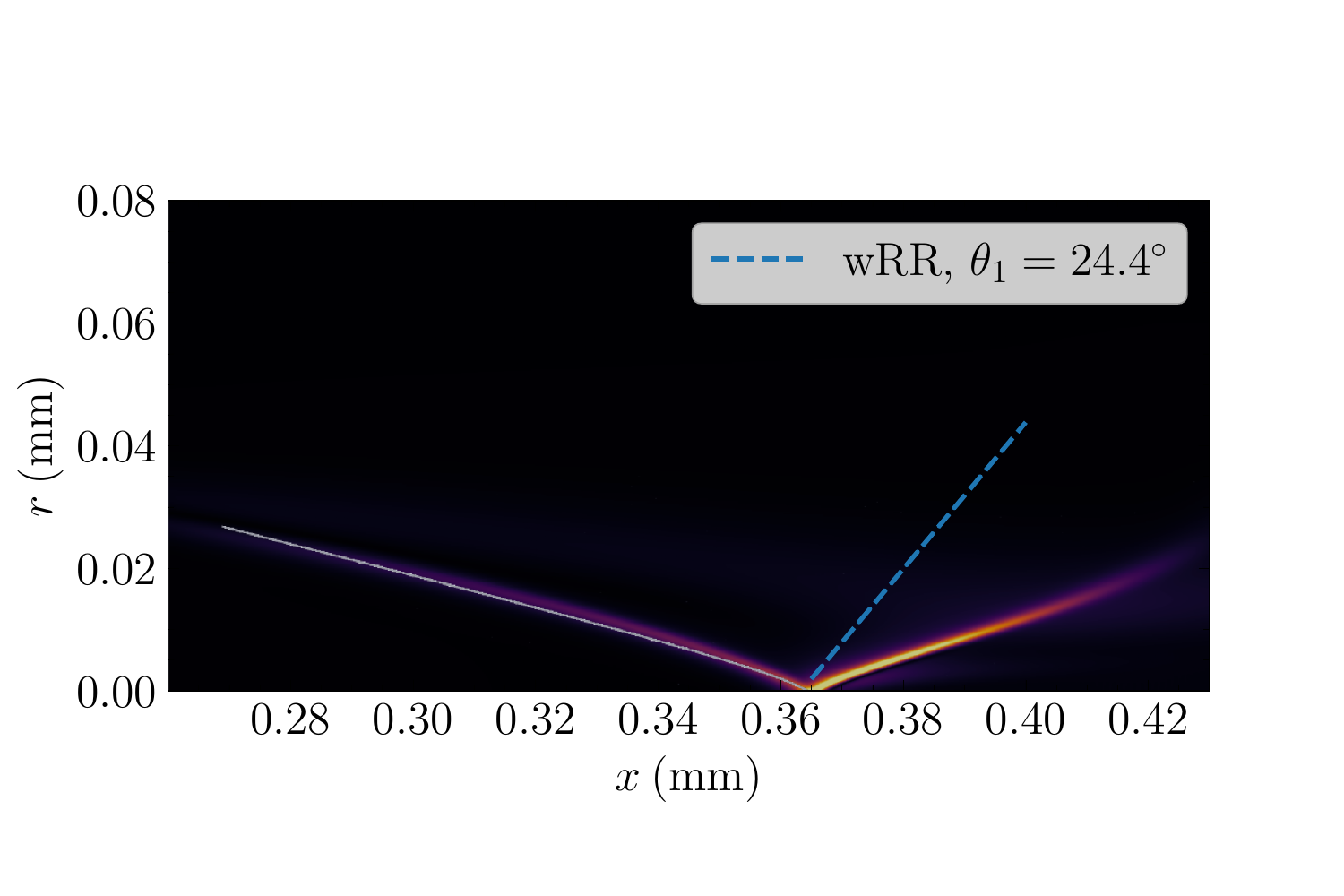}
    \DIFaddendFL \caption{Numerical schlieren image of \DIFaddbegin w-2SC \DIFaddend superimposed with Guderley's analogy fit ($m = 0.9$), where $m$ is the variable to be determined, for the shape of the incident oblique (conical) shock, as well as the reflected shock angle (blue dashed line), which is computed using inviscid two-shock theory \DIFaddbegin for planar weak regular reflection (wRR) \DIFaddend}
    \label{fig:guderley}
\end{figure}

The shape of the incident shock is fitted to a family of generalised hyperbolas \cite{hornung2001oblique},
\begin{equation}
\label{eqn:3}
\left(\frac{x_{0}-x}{a}\right)^{1 / m}-\left(\frac{r}{b}\right)^{1 / m}= 1
\end{equation}
where $a$ and $b$ are the semi-axes of the hyperbola, $x_0$ is the streamwise location of its centre, and $m$ is a positive real number.  All parameters are fixed based on the numerical solution except $m$. $x_0$ is known through the slope of the inward-facing conical shock, and $a$ and $b$ can be found by taking the difference between the computational solution and its asymptote, leaving only $m$ as the variable to be determined. Here, $m=0.9$ is utilised as it provides the best fit for the numerical results (Fig.~\ref{fig:guderley}). More details of this model can be found in Refs~\cite{hornung2001oblique,shoev2019numerical}.
This demonstrates that the analogy also holds in the case of highly viscous overexpanded microjets, where the method of generalised hyperbolas suffices to predict the shape and curvature of the incident oblique (conical) shock.
\section{Conclusion}
The viscous effects on centreline shock reflection occurring within an overexpanded axisymmetric microjet has been investigated numerically. An MR $\rightarrow$ RR transition has been observed within the axisymmetric nozzle geometry for the first time, supporting recent numerical results \cite{shoev2019numerical}. This evinces that there exists a limiting Reynolds number where the RR wave configuration can appear. 

Moreover, the grid resolution achieved in the present work far exceeds prior studies, thereby confirming the possibility of RR formation within the highly viscous regime of axisymmetric internal flows, which would otherwise be impossible in the inviscid regime. It has also been shown that the transition occurs within the dual-solution domain, which suggests that this behaviour is analogous to dual-solution domain transition in the case of planar shock reflection, according to von Neumann and detachment criterion. 

In order to further elucidate the effect of viscosity on centreline shock reflection, the analogy to the Guderley's singularity has also been employed to represent the shape of the incident oblique (conical) shock from a mathematical standpoint. The results demonstrate that despite the highly viscous nature of the nozzle flowfield, the behaviour of the shock structure is still well-captured by the analogy, validating its use in axisymmetric nozzle flows.

\backmatter

\bmhead{Acknowledgments}
The computations were conducted with high-performance computing resources provided by the National Computational Infrastructure (NCI) (grant xx52), which is supported by the Australian Government. We also acknowledge the support provided by Institute of Fluids Science (IFS), Tohoku University in the General Collaborative Research Project (J22I048).

\bmhead{Data Availability Statement}
The datasets generated during and/or analysed during the current study are available from the corresponding author on reasonable request.






\bibliography{sn-bibliography}


\end{document}